\documentclass[aip,reprint,amsmath,amssymb]{revtex4-1}
\usepackage{graphics}
\usepackage{subfigure}
\usepackage{epsfig}
\usepackage{epsf,epic}
\usepackage{color}
\usepackage{marvosym}
\usepackage{subfigure}
\usepackage{amsmath}
\usepackage{amssymb}
\usepackage{amsfonts}
\usepackage{wrapfig}
\usepackage{pstricks}
\usepackage{multirow}
\usepackage{bm}
\usepackage{dcolumn}
\newcommand{\etal}{\textit{et al.\ }}

\newcommand{\degreeC}{\ensuremath{^\circ \rm C}}

\begin{document}
\title{First-principles study of point defects in LiGaO$_2$}

\author{Adisak Boonchun} 
\author{Klichchupong Dabsamut}
\affiliation{Department of Physics, Faculty of Science. Kasetsart University, Bangkok 10900 Thailand}
\affiliation{Thailand Center of Excellence in Physics, Commission on the Higher Education, Bangkok 10400, Thailand}
\author{Walter R. L. Lambrecht}
\email{walter.lambrecht@case.edu}
\affiliation{Department of Physics, Case Western Reserve University, 10900 Euclid Avenue, Cleveland, Ohio 44106-7079, USA}

\begin{abstract}
  The native point defects are studied in LiGaO$_2$ using hybrid functional
  calculations. We find that the relative energy of formation of the
  cation vacancies and the cation antisite defects depends strongly on the
  chemical potential conditions. The lowest energy defect is found to be
  the Ga$_\mathrm{Li}^{2+}$ donor. It is compensated mostly
  by $V_\mathrm{Li}^{-1}$ and in part by Li$_\mathrm{Ga}^{-2}$ in the
  more Li-rich conditions. The equilibrium carrier concentrations are found to
  be negligible because the Fermi level is pinned deep in the gap and
  this is consistent with insulating behavior in pure LiGaO$_2$. 
  The $V_\mathrm{Ga}$ has high energy under all reasonable conditions. 
 Both the Ga$_\mathrm{Li}$ and the $V_\mathrm{O}$
  are found to be negative $U$ centers with deep $2+/0$  transition levels.
\end{abstract}

\maketitle
\section{Introduction}
Recently, there has been an interest in  ultra-wide-band-gap semiconductors
such as $\beta$-Ga$_2$O$_3$ because of their potential in pushing
high-power transitions to the next level of performance.\cite{Sasaki13,Green16} An important
figure of merit for such applications is the breakdown field and the
latter is directly correlated with the band gap.  Here we draw attention to an
even higher band gap material, LiGaO$_2$.
LiGaO$_2$ has a wurtzite-derived crystal structure\cite{Marezio65,Ishii98} and band gap of $\sim$5.3-5.6 eV (at room temperature)
based on optical absorption\cite{Wolan98,Johnson2011,Ohkubo2002,Chen14} but potentially even as large as 6.25 eV (at $T=0$) based on quasi-particle self-consistent (QS) $GW$ calculations,\cite{Boonchun11SPIE} with $G$ the one-particle
Green's function and $W$ the screened Coulomb potential. It can be thought of as a I-III-VI$_2$ ternary analog of wurtzite ZnO, in which each group II Zn atom is replaced
by either a group-I Li or a group III-Ga in a specific ordered pattern
with the $Pbn2_1$ spacegroup. In this structure the octetrule is satisfied
because each O is surrounded tetrahedrally by two Li and two Ga. 
The prototype for this crystal structure is
$\beta$-NaFeO$_2$. LiGaO$_2$ can be grown in bulk form by
the Czochralsky method\cite{Marezio65} and because of its good lattice match
has been explored as a substrate for GaN. It can also be
grown by epitaxial methods
on ZnO and vice versa. Mixed ZnO-LiGaO$_2$ alloys have been reported.\cite{Omata08,Omata11} 
It has been considered for piezoelectric properties,\cite{Nanamatsu72,Gupta76,Boonchun2010} and is naturally considered as a wide gap 
insulator. However, Boonchun  and Lambrecht \cite{Boonchun11} suggested
it might be
worthwhile considering as a semiconductor electronic material and
showed in particular that it could possibly be n-type doped by Ge.
That study only used the 16 atom primitive unit cell of LiGaO$_2$
and thus considered rather high (25 \%) Ge$_\mathrm{Ga}$ doping or Mg$_\mathrm{Li}$ doping. It did not study the site competition or native defect compensation
issues. Here we study the native point defects by means of hybrid functional
supercell calculations.

\section{Computational Method}
Our study is based on density functional calculations using
the Heyd-Scuseria-Ernzerhof (HSE) hybrid functional.\cite{HSE03,HSE06} The
calculations are performed using the Vienna Ab-Initio Simulation Package (VASP). \cite{VASP,KresseVasp1}
The electron ion interactions are described by means of the
Projector Augmented Wave (PAW) method.\cite{Blochl94,KresseVasp3} We use a
well-converged energy cut-off of 500 eV for the projector
augmented plane waves. We performed the calculations with
a supercell size of 128 atoms (which corresponds to $2\times2\times2$
the primitive unit cell) and
a single {\bf k}-point shifted away from $\Gamma$
is employed for the Brillouin zone integration. The valence
configurations used were $2s^1$ for Li,
$3d^{10}4s^24p^1$ for Ga and $2s^22p^4$ for O. In the HSE functional,
the Coulomb potential
in the exchange energy is divided into short-range and long-range
parts with a screening length of
10 \AA\   and only the short-range part of the exact Hartree-Fock non-local
exchange is included by mixing it with the generalized gradient
Perdew-Burke-Enrzerhof (PBE) potential with a mixing fraction $\alpha=0.25$.
The band gap obtained in this way ($E_g=5.10$ eV) is still slightly lower
than the experimental value. 

\section{Results}
The energy of formation of the defect $D^q$ in charge state $q$
is given by
\begin{equation}
  \begin{split}
    E_f(D^q)=E_{tot}(C:D^q)-E_{tot}(C)-\sum_i\Delta n_i \mu_i \\
    +q(\epsilon_v+\epsilon_F+V_{align})+E_{cor}
  \end{split}
\end{equation}
where $E_{tot}(C:D^q)$ is the total energy of the supercell containing the
defect and $E_{tot}(C)$ is the total energy of the perfect crystal supercell.
The chemical potentials $\mu_i$ represent the energy for adding or removing
atoms from the crystal to a reservoir in the process of making the defect.
The $\Delta n_i$ is the change in number of atoms of species $i$.
Likewise the chemical potential of the electron determining its charge
state is $\epsilon_F+\epsilon_v+V_{align}$ with $\epsilon_v$ the energy
of an electron at the valence band maximum (VBM) relative
to the average electrostatic potential in bulk and $\epsilon_F$ the Fermi energy in the gap measured from the VBM. 
The alignment potential $V_{align}$ represents the alignment of 
 the average electrostatic potential 
in the supercell far away from the defect relative to that in the bulk.
This is calculated using the
Freysoldt \etal approach.\cite{Freysoldt09,Freysoldt14}
The final term is the image charge
correction term which corrects for the Madelung energy of the periodic
array of net defect point charges in the uniform background that is added
to ensure overall charge neutrality when considering a locally charged
defect state. It is closely related to the alignment potential and
including these corrections allows one to extrapolate the energy of formation
to the dilute limit of an infinitely large supercell. 

\begin{figure}
  \includegraphics[width=7cm]{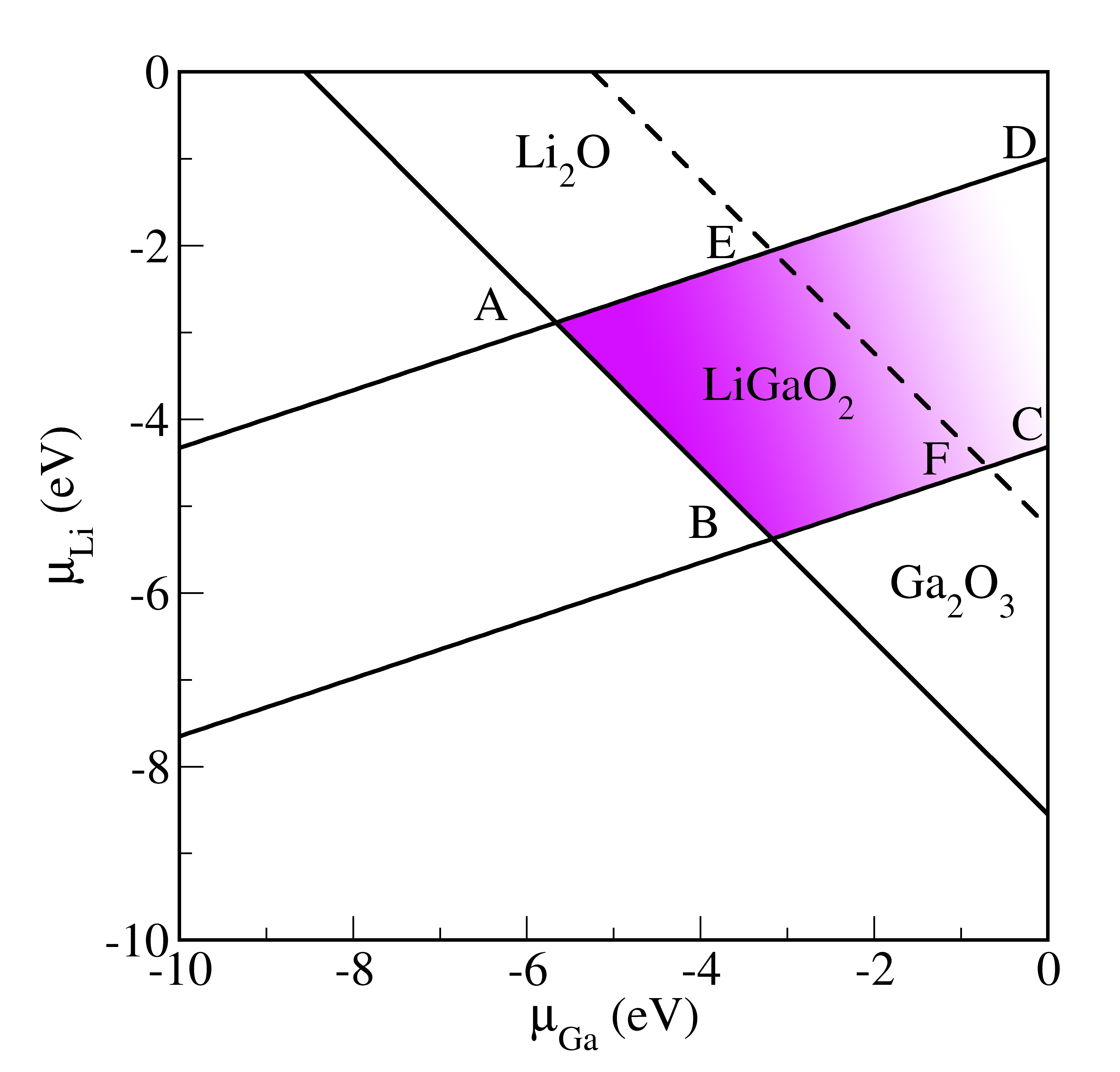}
  \caption{Chemical potential phase diagram, showing the region of stability of
    LiGaO$_2$. Note that these are excess chemical potentials indicated
    by $\tilde\mu$ in the text. \label{figchempot}}
\end{figure}
The chemical potentials $\mu_i=\mu_i^0+\tilde \mu_i$, where $\mu_i^0$ are the
chemical potentials of each species in its reference state, namely
the phase it occurs in at standard pressure and room temperature, and
$\tilde\mu_i$ are the excess chemical potentials. The latter 
are viewed as  a tunable
parameter reflecting the growth conditions but must obey certain
restrictions based on thermodynamic equilibrium.
These include
\begin{equation}
  \tilde\mu_\mathrm{Li}+\tilde\mu_\mathrm{Ga}+2\tilde\mu_{O}=\tilde\mu_{\mathrm{LiGaO}_2}
\end{equation}
where $\tilde\mu_{\mathrm{LiGaO}_2}$ is the energy of formation of
LiGaO$_2$, which we calculated to be $-8.55$ eV.
Each of the excess chemical potentials $\tilde\mu_i\le0$
on the left must be less than zero in order to avoid precipitation of the
bulk elements Li and Ga or evolving O$_2$ gas. For example,
$\mu_\mathrm{Li}^0$ corresponds to metallic body-centered-cubic Li and
thus $\tilde\mu_\mathrm{Li}=0$
corresponds to the assumption that the crystal with the defect is in equilibrium
with bulk metallic Li as reservoir. Similarly $\tilde\mu_\mathrm{Ga}=0$ corresponds to equilibrium with metallic bulk Ga and $\tilde\mu_\mathrm{O}$
corresponds to O in the O$_2$ molecule. However, we need to also
consider further restrictions imposed by competing binary compounds
Ga$_2$O$_3$ and Li$_2$O.
\begin{eqnarray}
  2\tilde\mu_\mathrm{Li}+\tilde\mu_\mathrm{O}&\le& \tilde\mu_{\mathrm{Li}_2\mathrm{O}},\nonumber \\
  2\tilde\mu_\mathrm{Ga}+3\tilde\mu_\mathrm{O}&\le&\tilde\mu_{\mathrm{Ga}_2\mathrm{O}_3}.
\end{eqnarray}
These restrictions determine the region of chemical potentials in which
LiGaO$_2$ is stable relative to the competing binaries and elements.
They are bounded by
\begin{eqnarray}
  \tilde\mu_\mathrm{Li}&\ge&\frac{1}{3}\tilde\mu_\mathrm{Ga}+[\tilde\mu_{\mathrm{LiGaO}_2}-\frac{2}{3}\tilde\mu_{\mathrm{Ga}_2\mathrm{O}_3}], \nonumber \\
    \tilde\mu_\mathrm{Li}&\le&\frac{1}{3}\tilde\mu_\mathrm{Ga}+\frac{1}{3}[2\tilde\mu_{\mathrm{Li}_2\mathrm{O}}-\tilde\mu_{\mathrm{LiGaO}_2}].
\end{eqnarray}
with
$\tilde\mu_{\mathrm{LiGaO}_2}-\frac{2}{3}\tilde\mu_{\mathrm{Ga}_2\mathrm{O}_3}=-4.32$ eV and $\frac{1}{3}[2\tilde\mu_{\mathrm{Li}_2\mathrm{O}}-\tilde\mu_{\mathrm{LiGaO}_2}]=-1.00$ eV.

It is represented in the phase diagram shown in Fig. \ref{figchempot}.
The points $A,B,C,D$ correspond respectively to (A) Li-rich, Ga-poor,
(B) Li-poor as well as relative Ga-poor, (C) Ga-rich, Li-poor and
(D) Ga-rich and Li-rich but O-poor.
The shading of the color is darker the higher the
chemical potential of O and the line $AB$ corresponds to
the O-rich limit $\tilde\mu_\mathrm{O}=0$.
In addition to the extreme chemical potential conditions (Ga-rich and Li-rich), we consider an intermediate oxygen chemical potential corresponding to a realistic growth condition during the annealing of LiGaO$_2$.
The oxygen chemical potential is a function of temperature and oxygen partial pressure, as described by Reuter \etal\cite{Reuter01}
	\begin{equation}
	\tilde{\mu}_\mathrm{O}(T,p)=\tilde{\mu}_\mathrm{O}(T,p_0)+\frac{1}{2}k_BT {\rm ln}(p/p_0),
	\end{equation}
	where $\tilde{\mu}_\mathrm{O}(T,p_0)$ is the oxygen chemical potential at the standard pressure $p_0=1$ atm, $k_B$ is Boltzmann's constant, and $T$ is the temperature in Kelvin. In the growth experiment of Ref. \onlinecite{Chen14}, the
	mixed $\rm Li_2CO_3$ and $\rm Ga_2O_3$ powders were compressed into tablets and
	then calcined at 1200 \degreeC\ for 20 h in air.\cite{Chen14} We therefore choose an annealing temperature of 1200\degreeC\ and an oxygen partial pressure of 0.21 atm which represents the ratio of oxygen gas in ambient environment. The growth conditions at annealing temperature of 1200 \degreeC\ and oxygen partial pressure of 0.21 atm is represented by the  dashed line $EF$ in Fig. 1.

\begin{figure}
  \includegraphics[width=10cm]{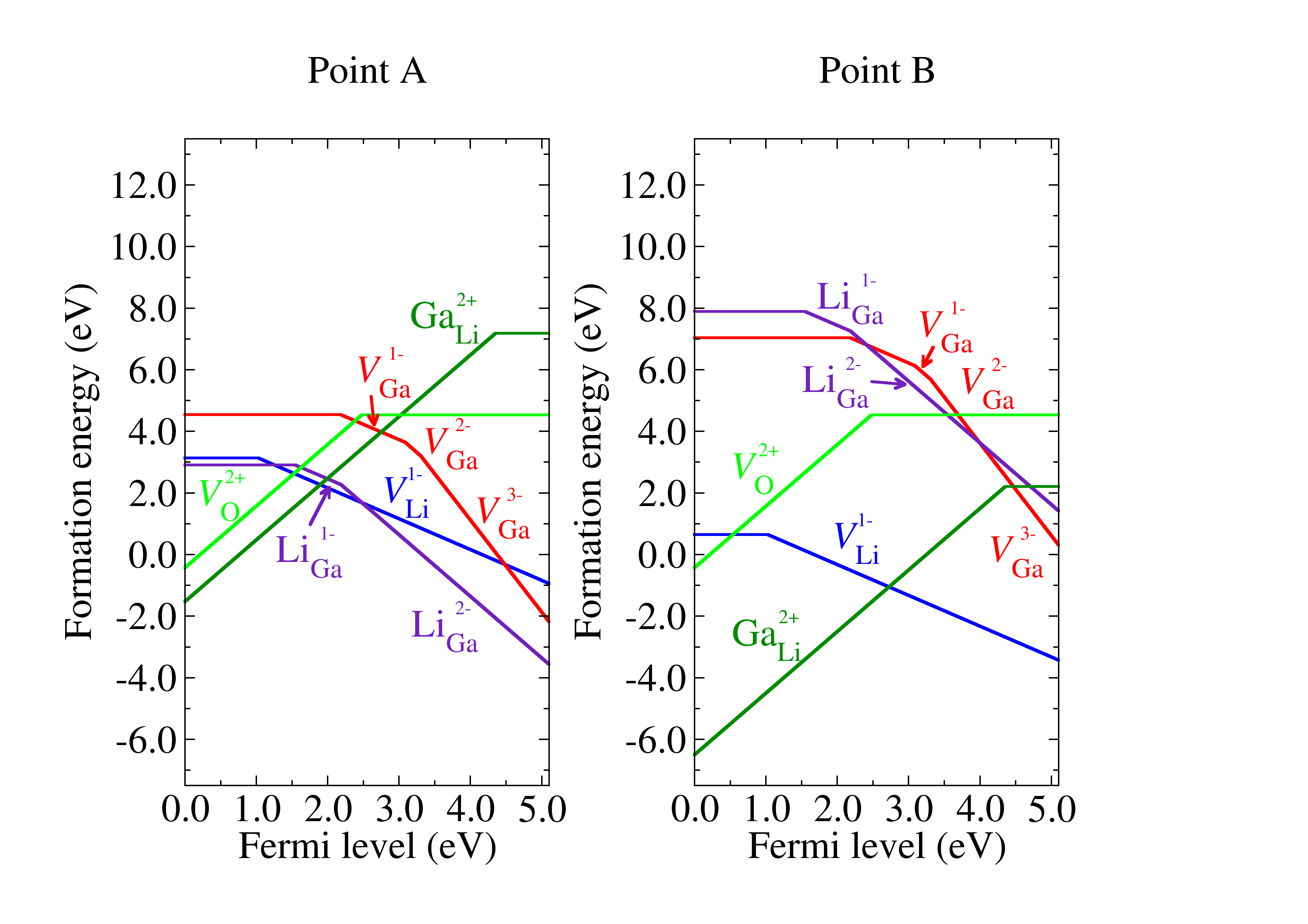}
  \includegraphics[width=10cm]{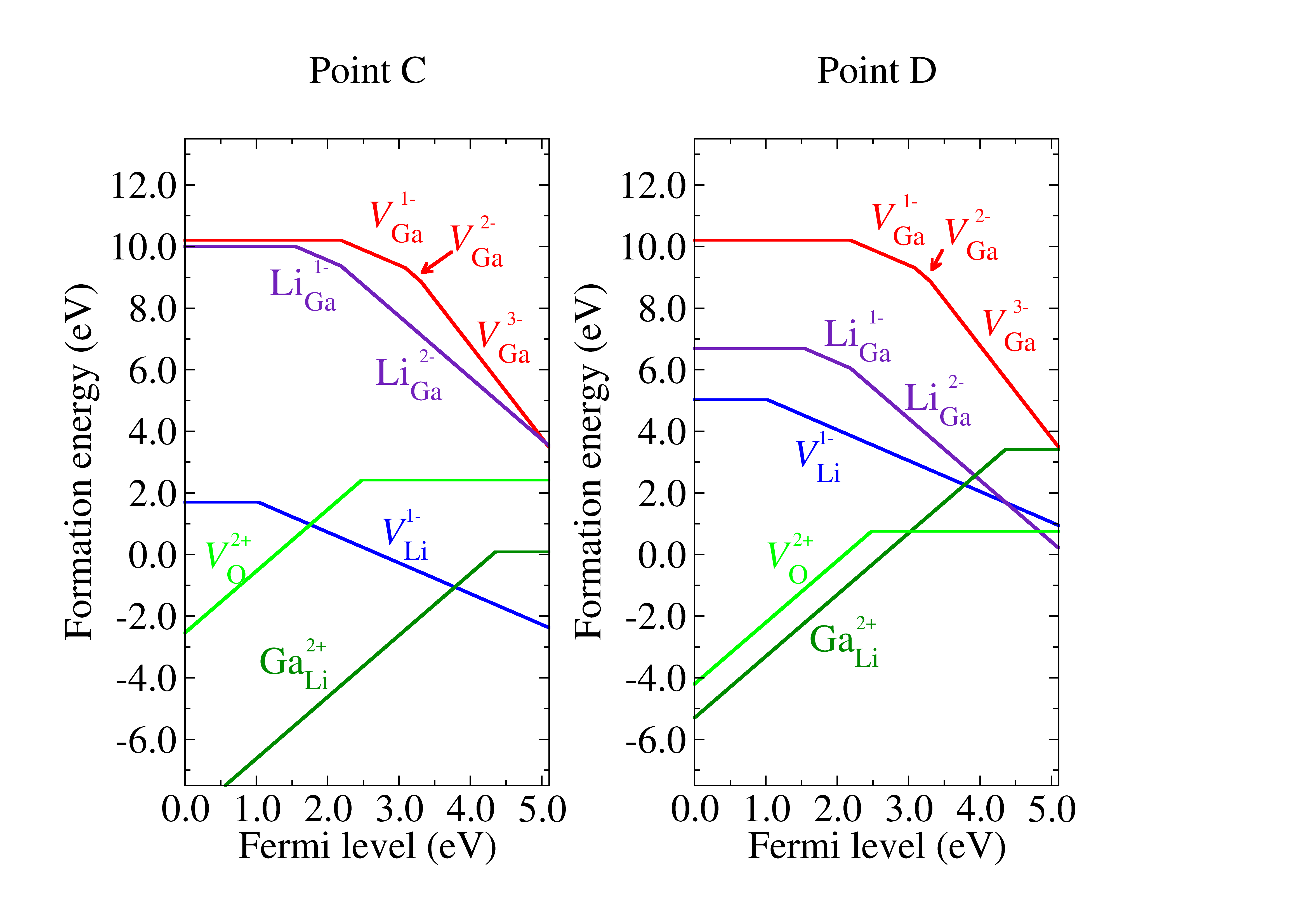}
    \includegraphics[width=10cm]{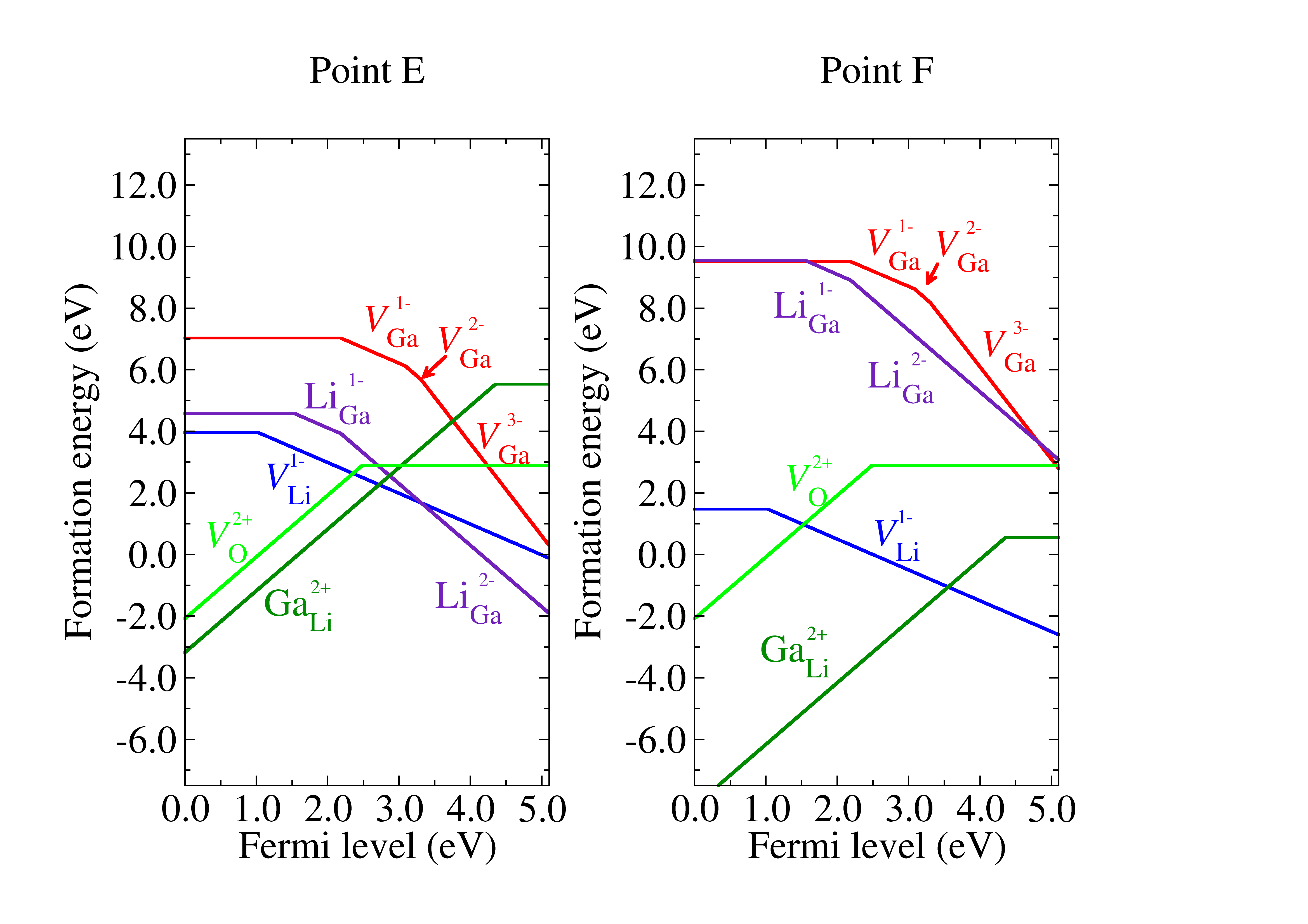}
  \caption{Energies of formation of various defects in LiGaO$_2$ for chemical potential conditions identified in Fig. \ref{figchempot}.\label{figefor}}
\end{figure}

The defects considered are the vacancies $V_\mathrm{Ga}$, $V_\mathrm{Li}$
and $V_\mathrm{O}$ and the antisites Li$_\mathrm{Ga}$ and Ga$_\mathrm{Li}$.  The effects of spin polarization were included for cases with unpaired electrons
in defect levels. 
Interstitial defects will be considered in the future but comparison
with II-IV-N$_2$ semiconductors suggest that they would be of high
energy.\cite{Skachkov16,Skachkov17}
The defect energies of formation are shown for the six chemical potential
points $A,B,C,D,E$ and $F$ in Fig. \ref{figefor}.

\begin{figure*}
	\includegraphics[width=18cm]{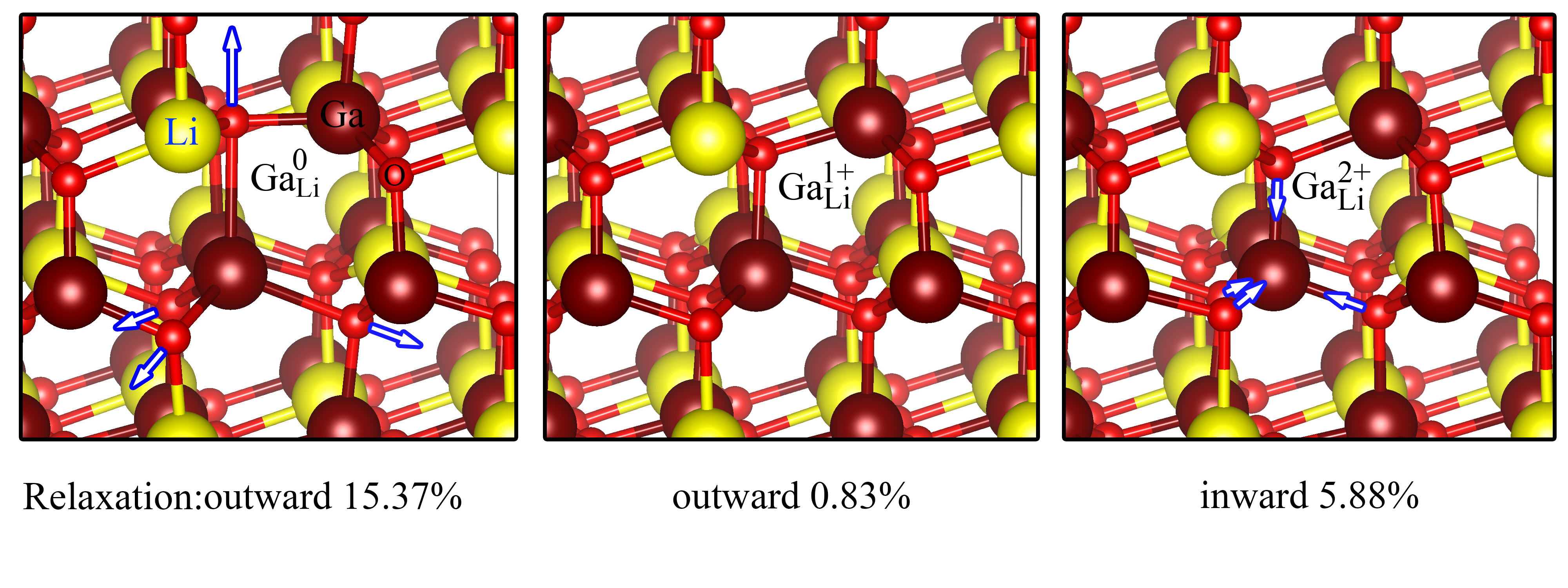}
	\caption{Structural relaxation for Ga$_\mathrm{Li}$ in different
          charge states.
        \label{figrelax}}
\end{figure*}

First we see that Ga$_\mathrm{Li}$ is the lowest energy defect for $\epsilon_F=0$
in all cases. It is a double donor, which is in the
$2+$ charge state over most of the gap. Still, it has a well-defined $2+/0$ transition making it a negative $U$ system.
In Fig. \ref{figrelax} we can see that while for the neutral charge
state, the O around Ga$_\mathrm{Li}$ move outward, they move
inward for the $2+$ charge state with an in-between outward relaxation
for the $1+$ state. The additional stabilization by outward motion of the
O when adding two electrons rather than one causes the negative $U$
behavior where the $1+$ charge state is never the lowest energy one
for any Fermi level position. It is thus not behaving like a simple
shallow donor, consistent with the relatively deep donor binding
energy of 0.74 eV below the conduction band minimum (CBM). We thus
do not expect it to be an effective n-type dopant. We can see that this
defect has negative energy of formation at $\epsilon_F=0$ in most cases.
This reflects that even in the most Ga-poor case, this defect is
hard to avoid because we cannot make the system poor enough in Ga without
reaching the stability limit imposed by Li$_2$O.  On the other hand,
a Fermi level $\epsilon_F=0$ is not expected to be realistic
as discussed later. 

The Li$_\mathrm{Ga}$ antisite on the other hand is a double acceptor which
can occur in $0,-1,-2$ charge states. It is the  lowest energy defect
in its $2-$ charge state near the CBM in cases $A$, $D$ and $E$. These are
the cases richest in Li.  

As for the vacancies,
$V_\mathrm{Li}$ occurs in $0,-1$ charge states, while $V_\mathrm{Ga}$ occurs
in $0,-1,-2,-3$ charge states.
We can see that $V_\mathrm{Ga}^0$ has a high energy of formation
in all cases. Although its negative charge states have significantly lower
energy for $\epsilon_F$ close to the CBM, it never becomes the lowest energy
defect and therefore does not play a role in determining the Fermi level.
The $V_\mathrm{Li}$ is more interesting. Although it has high energy
in the Li-rich case $D$ (which is somewhat unrealistic and O-poor)
it has low energy in the Li-poor cases, $B,C,F$. Even in case $E$,
its intersection with the Ga$_\mathrm{Li}^{2+}$ occurs close to that of the
intersection of the latter with Li$_\mathrm{Ga}^{2-}$.
We thus expect that both these acceptors may play a role in
compensating the Ga$_\mathrm{Li}^{2+}$.

\begin{figure*}
	\includegraphics[width=18cm]{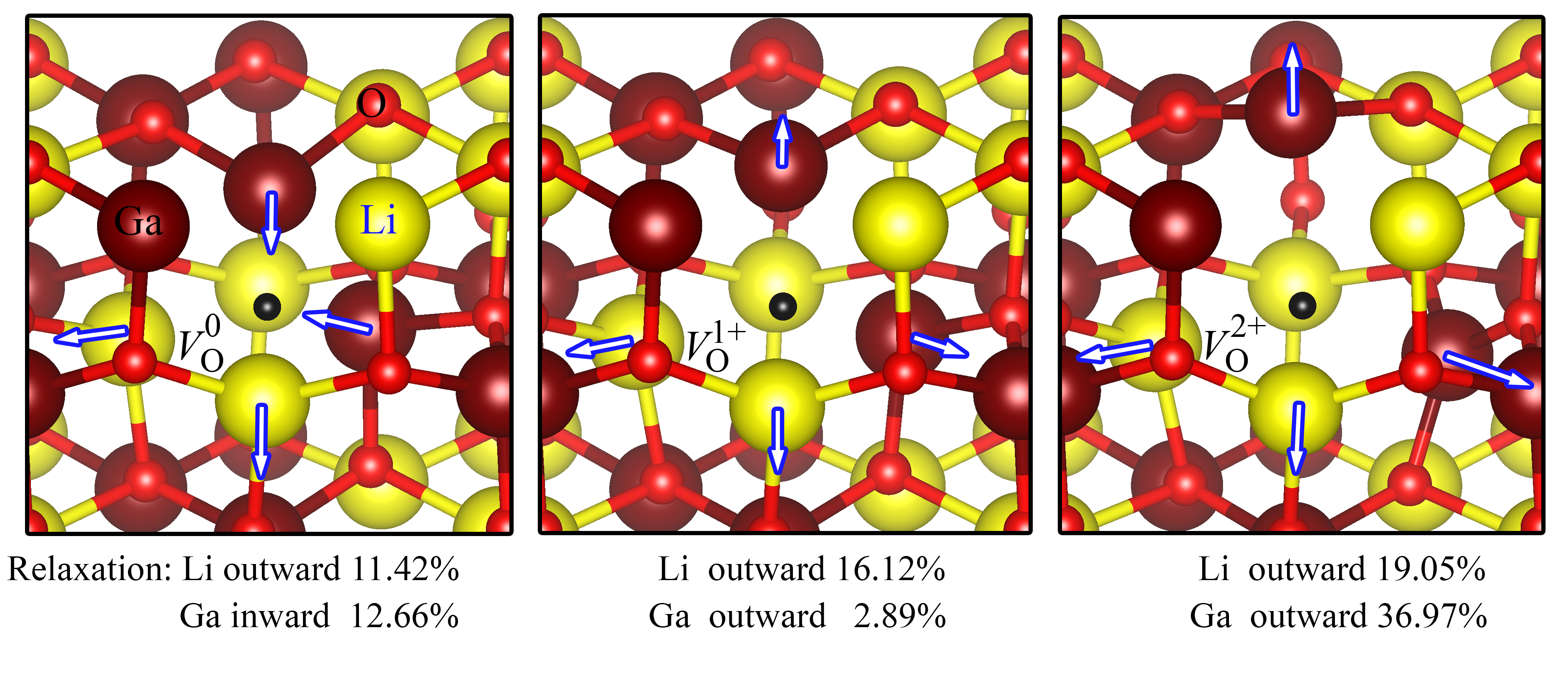}
	\caption{Structural relaxation for $V_\mathrm{O2}$ in different
          charge states.
        \label{figvo}}
\end{figure*}

Turning now to the O-vacancies, 
there are two non-equivalent sites for the oxygen in LiGaO$_2$:
on top of Li (O$_1$) or on top of Ga (O$_2$). We find
that both $V_\mathrm{O1}$ and $V_\mathrm{O2}$ are only stable in
the neutral and $2+$ charge states (with $V_\mathrm{O2}$ slightly lower
in energy than $V_\mathrm{O1}$)
with the transition level (2+/0) at 2.48 eV above the VBM or 2.62 eV below
the CBM. This is a quite deep donor level and indicates that
the vacancy is also a negative $U$ center. In Fig. \ref{figvo}
one can see that also in this case the relaxations are strongly charge-state dependent. This figure shows the relaxations near a $V_\mathrm{O2}$ but
similar results hold for $V_\mathrm{O1}$.
In the neutral charge state, the Ga move inward, while the Li move
outward. In the $2+$ state both move strongly outward. 
This is similar to the
$V_\mathrm{O}$ in ZnO\cite{Boonchun11,Boonchun13}
although the level is here even deeper and close to mid gap. 
We find that the $V_\mathrm{O}^{2+}$
energy of formation is negative for Fermi levels close to the VBM
for points $C,D,E,F$. They become positive for the O-rich limits ($A$, $B$). 
Its energy of formation is always higher than that of the Ga$_\mathrm{Li}^{2+}$
and thus it is not expected to play a significant role in the
charge balance. 


Using the charge neutrality condition between free electron
concentration $n_e(T,\epsilon_F)$, free hole  concentration $n_h(T,\epsilon_F)$
and the various defect concentrations,
\begin{equation}
  c(D^q; T,\epsilon_F)=N_Dg(q)e^{-(E_f(D^q,\epsilon_F=0)+q\epsilon_F)/k_BT}
\end{equation}
where $N_D$ is the number of available sites per cm$^{3}$ and $g(q)$ a
degeneracy factor depending on the charge state, we can find the equilibrium
Fermi level and the defect concentrations for a given temperature
following the procedure of Ref.\onlinecite{Skachkov16}. For the electron
and hole concentrations we use a parabolic band with effective density
of states masses $m_e^*\approx0.4$ and $m_h^*\approx1.8$ (as obtained
from the calculate hybrid functional band structure and averaging over
directons.)
For a temperature of $T=1500$ K close to the growth temperature,
we find that under chemical potential conditions $C$,  the equilibrium
Fermi level is $\epsilon_F=3.815$ eV, close to the intersection
of the $V_\mathrm{Li}^{1-}$ and Ga$_\mathrm{Li}^{2+}$. The electron
concentration  $n_e=6\times 10^{13}$ cm$^{-3}$ but  the
$[V_\mathrm{Li}^{-1}]=2[\mathrm{Ga}_\mathrm{Li}^{2+}]=1.0\times10^{26}$ cm$^{-3}$
are unrealistically high. This is related to the energies of formation
of the main defects Ga$_\mathrm{Li}$ and $V_\mathrm{Li}$
being negative for the equilibrium Fermi level.
For point $E$, the equilibrium Fermi level position is closer to mid gap, 
$\epsilon_F=2.75$ eV with $[\mathrm{Ga}_\mathrm{Li}^{2+}]=3.7\times10^{14}$,
$[V_\mathrm{Li}^{-1}]=7.22\times10^{14}$ and $[\mathrm{Li}_\mathrm{Ga}^{2-}]=1\times10^{13}$ cm$^{-3}$, $[V_\mathrm{O}^{2+}]=5\times10^{12}$ cm$^{-3}$.
So, in this case the concentrations of defects
  are much smaller and the Ga$_\mathrm{Li}^{2+}$ is still mostly 
  compensated by $V_\mathrm{Li}^{-1}$ but partially also by Li$_\mathrm{Ga}^{2-}$.
  The electron concentration at $n_e=3.2\times10^{12}$ cm$^{-3}$ is then
  only slightly higher than the hole concentration
  $n_h=1.6\times10^{11}$ cm$^{-3}$
  but both free carrier concentrations are in fact negligible under
  both chemical potential conditions considered.
  Even under the most Ga-poor conditions (point $A$), Ga$_\mathrm{Li}^{2+}$
  is the dominant defect and is compensated mostly by $V_\mathrm{Li}^{1-}$.
  In this case, $\epsilon_F=1.92$ eV is closest to the VBM
  and the material would then be slightly $p$-type with
  $n_h=9.7\times10^{13}$ cm$^{-3}$.

\begin{table}[h]
	\caption{
		Transition levels $\varepsilon(q,q')$ in eV relative to the VBM}	
	\label{tablevels}
	\begin{tabular*}{0.4\textwidth}{@{\extracolsep{\fill}}ccc}  \hline
		Defect & $q,q'$ & $\varepsilon(q,q')$  \\ \hline
		$V_{\rm Li}$&(0/1-)&1.0270 \\
$V_{\rm Ga}$&	(0/1-)&	2.1843\\
&(1-/2-) & 3.0899\\
&(2-/3-)	& 3.3088\\
$\rm Li_{Ga}$&	(0/1-)	&1.5464\\
&(1-/2-)	&2.1855\\
$\rm Ga_{\rm Li}$&	(0/2+)	&4.3569\\
$V_{\rm O}$&	(0/2+)&	2.4831\\ \hline
	\end{tabular*}
\end{table}

  It is instructive to compare the defect physics in this system
  to that in II-IV-N$_2$ semiconductors
  like ZnGeN$_2$,\cite{Skachkov16}. The similarity is that in both cases, 
  the antisites play a crucial role.
However, the dependence on chemical potentials of the elements is more important here because a wider region of stability occurs. Furthermore the Ga$_\mathrm{Li}$ antisite is here not a shallow but a deep donor and is thus not expected
to lead to unintentional n-type doping. This is consistent with the
insulating behavior of LiGaO$_2$. However, it does not exclude the possibility
of n-type doping by Si or Ge or Sn which will be studied separately. 

\begin{figure}
  \includegraphics[width=8cm]{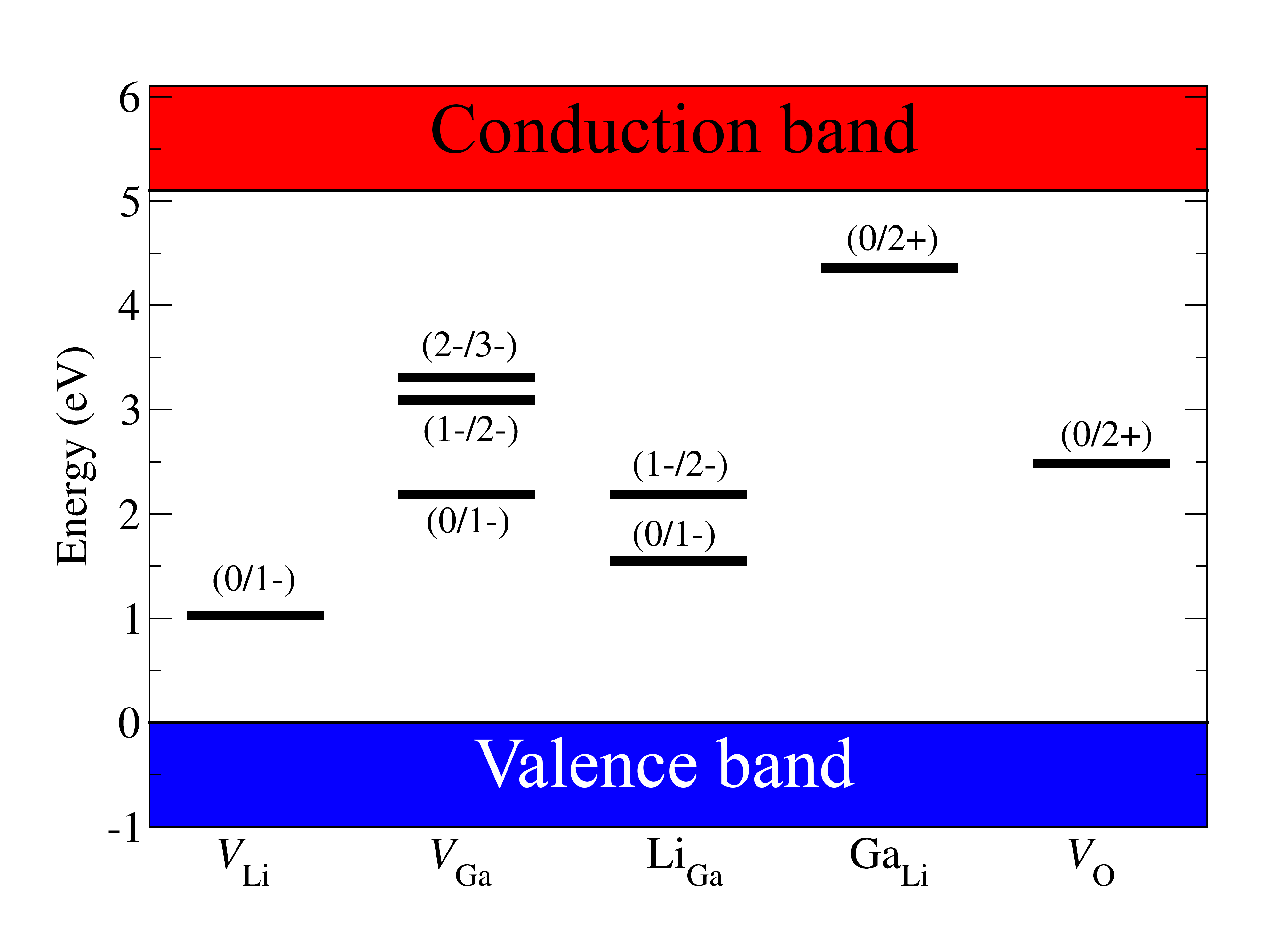}
  \caption{Defect transition levels in LiGaO$_2$.\label{figlevels}}
\end{figure}
The main defect transition levels in the gap are summarized in Table
\ref{tablevels} and in Fig. \ref{figlevels}.

\section{Conclusions}
In this paper we have studied the native defects in LiGaO$_2$. We find that
the relative energy of formation of vacancies and antisites depends strongly
on the chemical potential conditions.
The Ga$_\mathrm{Li}$ antisite is a dominant donor defect. 
However,
it has a rather deep  $2+/0$ donor level
and is a negative $U$ center. It is thus
not expected to lead to significant n-type doping. 
It furthermore becomes
compensated mostly by $V_\mathrm{Li}^{1-}$ and
in part by Li$_\mathrm{Ga}^{2-}$ depending
on how rich the system is in Li. The $V_\mathrm{O}$ is found
to be an even deeper double donor negative $U$ center. 
The defect transition levels are all
relatively deep in to the gap with no truly shallow levels.

\acknowledgements{The work at CWRU was supported by
  the U.S. National Science Foundation under grant No. 1755479. 
  The work at Kasetsart was supported by Kasetsart University Research and Development Institute (KURDI).}

\bibliography{Bib/dft,Bib/ligao2,Bib/defects,Bib/ga2o3}
\end{document}